\documentstyle[11pt,fleqn]{article}
\input{psfig.tex} %\parindent=0 pt
\title{K-essential Phantom Energy: Doomsday around the Corner? Revisited} % Declares the title.
\author{A. Kwang-Hua CHU \thanks{Permanent Address : P.O. Box 30-15, Shanghai 200030, PR China
  }}  %address%\cite{PKU:1999}
\date{Department of Physics, Xinjiang University, 14 Shengli Road,\\
Urumqi 830046, PR China}
\begin{document}           % End of preamble and beginning of text.
\maketitle
\begin{abstract}
We generalize some of those results reported by  Gonz\'{a}lez-D\'{i}az
 by further tuning the parameter ($\beta$) which is
closely related to
the canonical kinetic term in $k$-essence formalism.   The scale factor
$a(t)$ could be negative and decreasing within a specific range of $\beta$ ($\equiv -1/\omega$,
$\omega$ : the equation-of-state parameter)
during the initial evolutional period. \newline

\end{abstract}
%\pacs{98.80.Cq}
%%\end{titlepage}      %%\twocolumn     %%\nopagebreak   %\oddsidemargin=1mm
\doublerulesep=6mm    %
\baselineskip=6mm
\bibliographystyle{plain}               %
\section{Introduction}
The standard (hot) big bang (SBB) theory is an extremely successful one,
and has been around for over 60 years, since Gamow originally proposed it [1-2].
Remarkably, for such a simple idea, it provides us with an understanding
of many of the basic features of our Universe. All that you require in the
cooking pot, are initial conditions of an expanding scale factor, gravity, plus
the standard particle physics we are used to, to provide the matter in the
Universe. However, as mentioned the hot big bang theory can successfully proceed only
if the initial conditions are very carefully chosen, and even then it only really
works at temperatures low enough, so that the underlying physics can be well
understood. The very early Universe is out of bounds, yet there is a hope that
accurate observations of the present state of the Universe may highlight the
types of process occurring during these early stages, providing an insight
on the nature of physical laws at energies which it would be inconceivable to
explore by other means. Another unresolved issue is the cause of the apparent
acceleration of the Universe today, as seen through the distribution of distant
Type Ia Supernovae. \newline
The physical properties of vacuum phantom energy are rather weird,
as they include violation of the dominant-energy condition, $P +
\rho <0$ (with the equation of state : $P=\omega \rho$ [3-4], $\omega <-1$;
$P$ is  the homogeneous dark energy pressure and $\rho$ is the energy density),
naive superluminal sound speed and increasing
vacuum-energy density. The latter property ultimately leads to the
emergence of a singularity usually referred to as big rip in a
finite time in future where both the scale factor and the
vacuum-energy density blow up [5-7]. For instance,  Gonz\'{a}lez-D\'{i}az
showed that a cosmological model with a singular big rip
at an arbitrary finite time in the future can be also obtained
when the scalar field satisfies equivalent phantom-energy
conditions in the case that it is equipped with non-canonical
kinetic energy [4] for models restricted by a Lagrangian of the
form ${\cal L} = K(\phi) q(y)$ [5]. He claimed that, one can play
with the arbitrary values of the prefactor $a_0$ for the scale
factor expression in Eq. (10) and those unboundedly small positive
values of $t$ which satisfy the observational constraint [5] (note
that in this model $\beta = -1/\omega_{\phi}=-1/\omega$) and the currently
observed cosmic acceleration rate [8] to check that such a set of
present observations [8] is compatible with unboundedly small
positive values of $t$ (the assumption is, if a quintessential
scalar field $\phi$ with constant equation of state $P_{\phi} =
\omega_{\phi} \rho_{\phi}=\omega \rho_{\phi}$ is considered, then phantom energy can
be introduced by allowing violation of dominant energy condition,
$P_{\phi} + \rho_{\phi} < 0$). Note that, in general k-essence is defined as a scalar field with
non-canonical kinetic energy, but usually the models are
restricted to the above Lagrangian form (${\cal L}$) [9]. Normally we use
units where $8\pi G/3$ = 1 [4].\newline
Gonz\'{a}lez-D\'{i}az proposed that, from the perfect-fluid
analogy, we have for the pressure and energy density of a generic
$k$-essence scalar field $\phi$ [9]
\begin{displaymath}
 P_{\phi} (y)=K(\phi) g(y)/y,  \hspace*{16mm} \rho_{\phi}
 (y)=-K(\phi) g'(y),
\end{displaymath}
where
\begin{displaymath}
  g(y)=B y^{\beta},
\end{displaymath}
  with $B$ and $\beta$ being given
constants such that $B >0$ and $0< \beta <1$ and the prime means
derivative with respect to $y$. Thus, the equation-of-state parameter reads
\begin{displaymath}
 \omega =\frac{- g(y))}{y g'(y)}.
\end{displaymath}
We set next the general form of the function $g(y)$
when we consider a phantom-energy $k$-essence field;
i.e., when we introduce the following two phantom energy
conditions: $K(\phi)<0$ and
$P_{\phi} (y)+ \rho_{\phi} (y)  <0$. The weak energy condition
$\rho_{\phi} (y)>0$ is presumed to be valid here so that $g'(y)>0$.
From the second phantom-energy condition, we then deduce
that $g(y) > yg'(y)$ (and $g(y)>0$). Therefore the function g(y)
should be an increasing concave function, that is
we must also set $d^2 g(y)/d y^2 < 0$.\newline
Following [5], we then specialize in the case of a spatially flat
Friedmann-Robertson-Walker spacetime with line element
$ds^2=-dt^2+a(t)^2 dr^2$, in which $a(t)$ is the scale factor. In
the case of a universe dominated by a $k$-essence phantom vacuum
energy, the Einstein field equations are then [4-5]
\begin{displaymath}
 3 H^2=\rho_{\phi} (y), \hspace*{12mm} 2 \dot{H}+\rho_{\phi} (y)+P_{\phi}
 (y)=0,
\end{displaymath}
with $H=\dot{a}/a$, the overhead dot meaning time derivative.
\newline
By combining the two expressions above and using the equation of
state he can obtain for the function $g(y)$ as given above
\begin{equation}
 3 H^2 =\frac{2 \dot{H} \beta}{1-\beta},
\end{equation}
and then the solutions (for his spatially flat case)
\begin{equation}
 a(t)=\frac{a_0}{(t-t_*)^{2\beta/[3(1-\beta)]}}, \hspace*{12mm}
 0<\beta<1,
\end{equation}
where $t_*$ is an arbitrary and positive constant (this positive
$t_*$ solution family does represent an accelerating universe if
$a(t)$ increases once $t$ increases [5]). Gonz\'{a}lez-D\'{i}az
presented the behavior of $a(t)$ in Fig. 1 (II) therein [5] and
made remarks : {\it of quite greater interest is the choice $t_*
> 0$ for which the universe (Fig. 1(II), solid line [5]) will
first expand to reach a big rip singularity at the arbitrary time
$t = t_*$ in the future, to thereafter steadily collapse to zero
at infinity; that is it matches the behaviour expected for current
quintessence models with $\omega < -1$. The potentially dramatic
difference is that whereas in quintessence models the time at
which the big rip will occur depends nearly inversely on the
absolute value of the state equation parameter, in the present
k-essence model the time $t_*$ is a rather arbitrary parameter}
[5].
\newline The present author, however, based on similar derivations
($1/y^2=\dot{\phi}^2/2$ [4-5]) could obtain Eq. (10) the same as
that in [5] but different behavior of $a(t)$ for $t<t_*$ region
with $\beta=1/2223$, $0.60$ and $297/299$ (compared to the (II) of Fig.1
in [5] which is for $0<\beta<1$). Our results are illustrated in
Fig. 1 with $\beta=1/2223, 0.60, 0.85, 297/299$, respectively
($t_*=2.0$). We can clearly observe that the curves of $\beta=1/2223,
0.60, 297/299$ (in $t<t_*$ region) are different from that claimed in
Fig. 1 (II) in [5]. Under these situations, values of $a(t)$
(which are negative, if we define
\begin{equation}
 B_p=\frac{2\beta}{3(1-\beta)}=\frac{-2/\omega}{3(1+\frac{1}{\omega})}=\frac{-2}{3(\omega+1)},
\end{equation}
we have
$B_p=1/3333, 1, 99$ for $\beta=1/2223, 0.60, 297/299$, respectively; then
(the scale factor)
\begin{equation}
 a(t) = \frac{a_0}{[(t-t_*)^{B_p}|]_{B_p=1/3333,1,99}}
\end{equation}
should be negative for $t<t_*$!) firstly decrease as (time) $t$
increases until $t \rightarrow t_*$. Here, $a_0$ is an arbitrary integration constant.
 \newline To be specific, considering $\beta=0.60$ and $297/299$ (or $\omega=-5/3$ and $-299/297$),
of which the role of dark energy
can be played by physical fields with positive energy
and negative pressure which violates the strong energy
condition $\rho + 3 P > 0 \, ( \omega  > -1/3 )$,
we thus obtain
\begin{equation}
 a(t)={a_0}{(t-t_*)^{-1}}, \hspace*{24mm} a(t)=\frac{a_0}{(t-t_*)^{99}}
\end{equation}
and there is no doubt that $a(t)$ should be negative once $t<t_*$.
These cases correspond to the
earlier collapsing ones (but not collapse to zero) instead of expansion. \newline
The $\beta=0.85$ curve (in
$t<t_*$ region), however, resembles that presented by
Gonz\'{a}lez-D\'{i}az in Fig. 1 (II) of [5]. It means, for some
cases of $\beta$, the universe will {\bf not } first expand to
reach a big rip singularity at the arbitrary time $t = t_*$ in the
future as Gonz\'{a}lez-D\'{i}az claimed in [5]. Cases of $a(t)$
being negative (in $t<t_*$ region) should be interpreted in
another way (at lease for cases of $\beta=1/2223,0.60,297/299$ where the
power : $B_p$ is, al least, an odd integer or the denominator of
$B_p$ is an odd integer while the numerator of $B_p$ is normalized
to be $1$.)! \newline In fact,  we can set
\begin{equation}
  B_p =\frac{n}{m}, \hspace*{24mm} n,m \not =0,
\end{equation}
where $n,m$ are positive integers. We already demonstrated cases of $B_p =1/3333, 1$, and $99$
which correspond to $(n,m)=(1,3333),(1,1),(99,1)$ or $\beta=1/2223,3/5,297/299$ (or $\omega=-2223,-5/3,-299/297$).
To let $B_p$ be a positive odd integer, we can select $m=1$ and $B_p=n$. It leads to
\begin{equation}
 \beta=\frac{3n}{3n+2}=\frac{3 B_p}{2+3 B_p}=\frac{-1}{\omega}.
\end{equation}
Under this choice, $a(t)$ will decrease (for $t<t_*$) following
\begin{equation}
 a(t)=\frac{a_0}{(t-t_*)^{B_p}}=\frac{a_0}{(t-t_*)^{n}}
\end{equation}
since $n$ is an odd integer ($t-t_* <0$)! For example, with $m=1$, $n=101$,
we have decreasing $a(t)$ for $\beta=303/305$! \newline
On the other hand, once we choose $n=1$, together with $m$ is a positive integer,
it then sets $B_p=1/m$ and
\begin{equation}
 \beta=\frac{3}{2m+3}=\frac{-1}{\omega}.
\end{equation}
 With this, as we have shown above $(n=1,m=3)$, $a(t)$ will decrease for $t<t_*$ following
\begin{equation}
 a(t)=\frac{a_0}{(t-t_*)^{B_p}}=\frac{a_0}{(t-t_*)^{1/m}}=\frac{a_0}{\sqrt[m]{t-t_*}}.
\end{equation}
One another example, for $\beta \sim 0$, is $m=1001$ ($B_p=1/1001$) or $\beta=3/2005$, which leads to
\begin{displaymath}
  a(t)=\frac{a_0}{\sqrt[1001]{t-t_*}}.
\end{displaymath}
To make a brief summary, for those cases considered in Eqs. (7) and (9), we have earlier
collapsing cases ($a(t)$ decreasing once $t< t_*$; a negative divergence!). However, as we have illustrated above,
cases which don't belong to the restriction of Eqs. (7) or (9), say, $n$ is an even integer
($m$ is still an odd integer), e.g., $\beta=0.85=17/20$ gives $n=34$ and $m=9$ or $B_p=34/9$,
will match the results claimed in [5] : the universe  will
first expand to reach a big rip singularity at the arbitrary time
$t = t_*$! At least, our results can also be extended to the cases considered in [3].
{\it Acknowledgements.} The author is partially supported by the Starting Funds of
XJU-2005-Scholars.

\newpage

\oddsidemargin=3mm

\pagestyle{myheadings}

\topmargin=-18mm

\textwidth=17cm \textheight=26cm
%\begin{document} %\psdraft         %wn0.s      &   un0.ps
\psfig{file=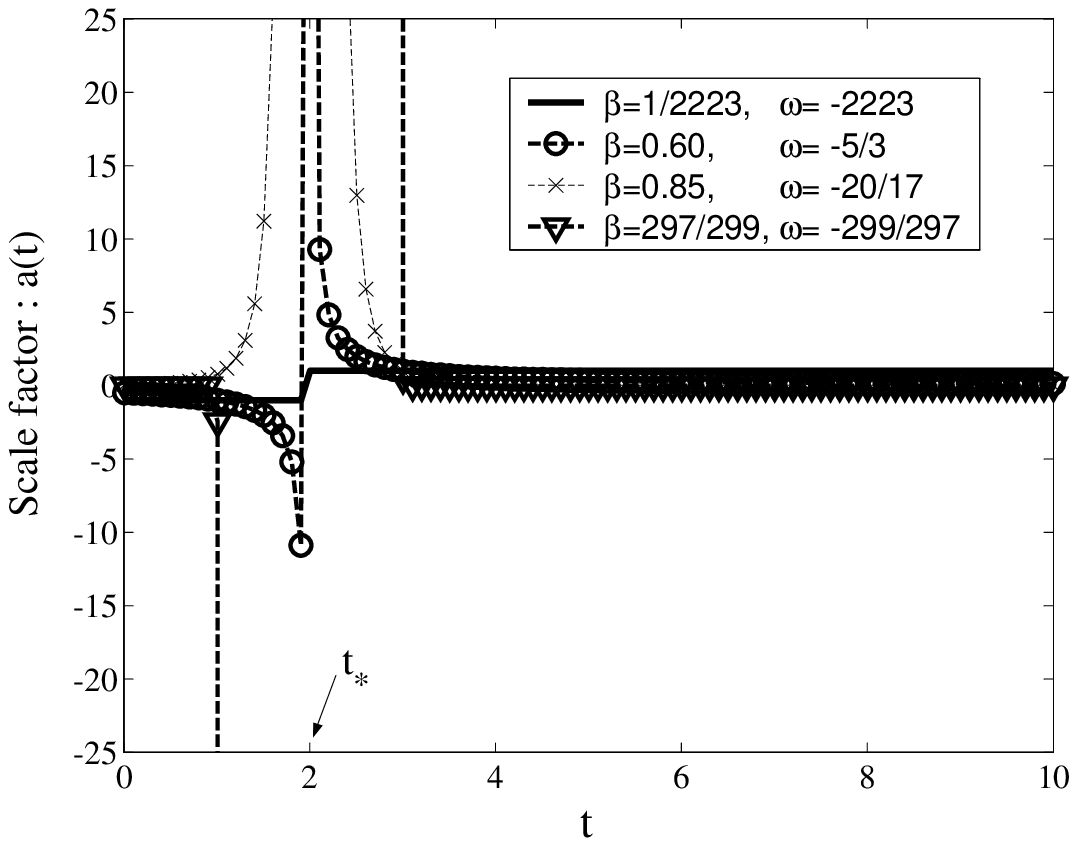,bbllx=0.1cm,bblly=10.5cm,bburx=12cm,bbury=23.8cm,rheight=9cm,rwidth=9cm,clip=}
%
%\vspace{2mm}
\begin{figure}[h]
\hspace*{3mm} Fig. 1 \hspace*{1mm} Evolution of the scale factor
$a(t)$ with cosmological time $t$ for a function $g(y)$ \newline
\hspace*{3mm} with the form given as $B y^{\beta}$. If we choose
that $t_*$ to be positive, then the constant $t_*$ (=2.0)
\newline \hspace*{3mm} becomes an arbitrary time in the future at
which the big rip takes place. All units in \newline \hspace*{3mm}
 the plot are also arbitrary. Cases of $\beta=1/2223,0.60,297/299$ are
different from that in  \newline \hspace*{3mm} Fig. 1 (II) in [5]
therein. $a(t) <0$ in $t<t_*$ region and $a(t)$ decreases as $t$
increases  \newline \hspace*{3mm} until $t \rightarrow t_*$.
The equation-of-state parameter $\omega=-1/\beta$.
%Schematic plot for the regular
%scattering and the disorder-influenced scattering. \newline
%\hspace*{7mm} Plane waves propagate along the $X$-direction.
%Binary encounters of $U_1$ and $U_3$ and their \newline
%\hspace*{7mm} departures  after head-on collisions ($U_2$ and
%$U_4$). Number densities $N_i$ are associated to $U_i$.
\end{figure}

\end{document}